\documentclass[times]{aastex631}

\usepackage{ulem,textcomp}

\begin{document}

\title{Spatiotemporal Variations of Temperature in Jupiter's Upper Atmosphere}

\author[0000-0002-5753-1262]{Kate Roberts}
\affiliation{Department of Astronomy, Boston University, Boston, MA, USA
}

\author[0000-0003-4481-9862]{Luke Moore}
\affiliation{Department of Astronomy, Boston University, Boston, MA, USA
}
\affiliation{Center for Space Physics, Boston University, Boston, MA, USA}

\author[0000-0002-4218-1191]{James O'Donoghue}
\affiliation{Department of Meteorology, University of Reading, Reading, UK
}

\author[0000-0001-5971-2633]{Henrik Melin}
\affiliation{Department of Physics, Mathematics \& Electrical Engineering, Northumbria University, Newcastle upon Tyne, UK
}

\author[0000-0003-3990-670X]{Tom Stallard}
\affiliation{Department of Physics, Mathematics \& Electrical Engineering, Northumbria University, Newcastle upon Tyne, UK
}

\author[0000-0001-5055-8115]{Katie L. Knowles}
\affiliation{Department of Physics, Mathematics \& Electrical Engineering, Northumbria University, Newcastle upon Tyne, UK
}

\author[0000-0002-6917-3458]{Carl Schmidt}
\affiliation{Department of Astronomy, Boston University, Boston, MA, USA
}
\affiliation{Center for Space Physics, Boston University, Boston, MA, USA}

\author[0000-0001-7339-9495]{Paola I. Tiranti}
\affiliation{Department of Physics, Mathematics \& Electrical Engineering, Northumbria University, Newcastle upon Tyne, UK
}

\begin{abstract}

Global temperatures in Jupiter's upper atmosphere are poorly constrained. Other than an \textit{in situ} measurement by the Galileo Probe, all temperature data come from remote sensing methods which primarily rely on emissions from H$_3^+$, the dominant molecular ion in giant planet ionospheres. While H$_3^+$ temperature serves as a proxy for thermospheric temperature under specific conditions, the available H$_3^+$ observations at Jupiter have limited spatial coverage and a wide range of reported temperatures that complicate analysis of atmospheric temperatures. We present high resolution H$_3^+$ temperature maps near local solar noon collected over three half-nights in 2022 and 2023. Pole-to-pole temperature structure is consistent across time spans of one month to one year. Median equatorial ($\pm$ 25$^{\circ}$ latitude) temperature across all three nights is 762 $\pm$ 43 K, with night-to-night differences of $<$75 K. Temperatures within the statistical locations of the northern and southern auroral ovals are 1200 $\pm$ 96 K and 1143 $\pm$ 120 K, respectively. A region $\sim$30 K cooler than its surroundings is found near 20$^{\circ}$ N, 90$^{\circ}$ W System III longitude, roughly coincident with a magnetic field anomaly, providing additional evidence for magnetic influence on Jupiter's upper atmosphere. Temperatures generally decrease smoothly from auroral to equatorial latitudes, consistent with the expected gradient if Jupiter’s non-auroral latitudes are heated primarily by dynamical redistribution of auroral energy.

\end{abstract}




\section{Introduction} \label{sec:intro}

Non-auroral temperatures in Solar System giant planet upper atmospheres are observed to be significantly hotter than anticipated based on solar EUV (extreme ultraviolet) heating rates. Possible non-solar sources of energy that may explain observed equatorial temperatures include non-auroral energetic particle precipitation, dissipation of upward-propagating gravity waves, and global redistribution of auroral energy (topics reviewed in \cite{Yelle2004}, Section 9.3.3). There are various complications associated with these additional energy sources though, each would leave a unique spatiotemporal signature in global temperatures if present. Upper atmospheric dynamics are controlled by temperature gradients, and strong constraints on temperature variations are essential to inform simulation energy inputs. Without them, predictions of giant planet general circulation and atmospheric evolution are limited.

At Jupiter, this temperature discrepancy was first noticed more than 50 years ago \citep{Strobel1973,Hubbard1972}. Subsequent studies have been mostly inconclusive in identifying the source(s) of additional non-solar heating. As summarized by \cite{Yelle2004}, early theoretical treatments of gravity waves at Jupiter were highly idealized, and could lead to a net heating or cooling in the thermosphere. A more recent treatment demonstrated that waves with properties consistent with those observed by Galileo and New Horizons can provide substantial upper atmospheric heating when a type of gravity wave energy dissipation, rovibrational damping, was incldued \citep{Lian2019}. There have been only a few general circulation models (GCMs) of Jupiter's upper atmosphere, and those too reach inconsistent conclusions regarding the heightened equatorial temperatures. For example, the Jupiter Thermospheric General Circulation Model (JTGCM) was able to dynamically heat the equatorial thermosphere to observed values when applying a parameterized auroral heating input \citep{Bougher2005}. However, non-auroral temperatures were too large when a low-latitude particle “drizzle" parameterization was also included, consistent with a lack of convincing evidence for such precipitation. On the other hand, subsequent azimuthally symmetric \citep{Smith2009} and fully 3-D \citep{Yates2020} Jupiter GCMs that also included magnetospheric coupling found that auroral energy was largely constrained to high latitude by the strong Coriolis forces which turn equatorward winds westward. 

Observed temperatures have similarly been too limited in scope to provide comprehensive constraints to date. An ionospheric hot spot was found above the Great Red Spot (GRS), possibly a signature of active wave heating \citep{ODonoghue2016}, but evidence for other such hot spots is sparse, and many observations do not see evidence of localized hot spots at the GRS at all \citep{Melin2024}. The closest “full-planet" coverage is a map in \cite{Lam1997} which was ahead of its time in terms of improved global observational coverage, but was unevenly sampled in longitude with low spectral resolution, complicating low-latitude temperature retrievals. Constraints on global temporal variations are once again limited. \cite{ODonoghue2021} contains two half-nights of observation covering roughly the same magnetic longitudes, but one night only covers the northern hemisphere. Other observations of Jupiter's upper atmosphere come from a variety of different instruments and span a wide range of temperatures across the planet ($\sim$500-1600 K), but are difficult to compare in any consistent basis due to limited spatial and/or temporal coverage \citep{Miller2020}. 

Here, we present new pole-to-pole observations of Jupiter's upper atmospheric temperature with $\sim$170$^{\circ}$ of longitudinal coverage, including 90$^{\circ}$ of direct overlap across three nights from 2022-2023 (\textbf{Section \ref{sec:obs}}). Results from these observations, all taken using the same observatory and instrument, follow from the same data reduction pipeline for consistency, as detailed in \textbf{Section \ref{sec:methods}}. The final maps shown in \textbf{Section \ref{sec:results}} reveal clear temperature trends, allowing for significantly improved constraints on spatiotemporal low-latitude temperature variations, as discussed in \textbf{Section \ref{sec:disc}}.

\section{Observations} \label{sec:obs}

We observed Jupiter using the Keck II Near InfraRed Spectrometer (NIRSPEC) \citep{McLean1998, Martin2018} on three half-nights: 15 December 2022, 22 November 2023, and 30 December 2023 UT. We used NIRSPEC’s high resolution mode with the KL-filter; this covers wavelengths from 2.134 to 4.228 microns over seven spectral orders (M = 21 - 27) with an echelle angle of 62.02$^{\circ}$ and cross-disperser angle of 33.56$^{\circ}$. The 0.288x24$''$ slit was paired with six coadds, each of approximately nine seconds, producing a high spectral resolution ($\lambda / \Delta \lambda$ = 25,000 - 30,000) spectrum. The instrument had a spectral plate scale of 0.098$''$/pixel and a spatial plate scale of 0.129$''$/pixel with a 2048$\times$2048 pixel detector while its slit-viewing camera (SCAM) \citep{Martin2016} had a plate scale of 0.157$''$/pixel with 256$\times$256 pixels and a shortpass filter covering 1.0 - 2.5 $\mu$m. The nights of observation coincided with the 47$^{th}$, 56$^{th}$, and 57$^{th}$ close approaches of Jupiter by the NASA orbiter Juno.

\begin{figure}
    \centering
    \includegraphics[width=0.9\textwidth]{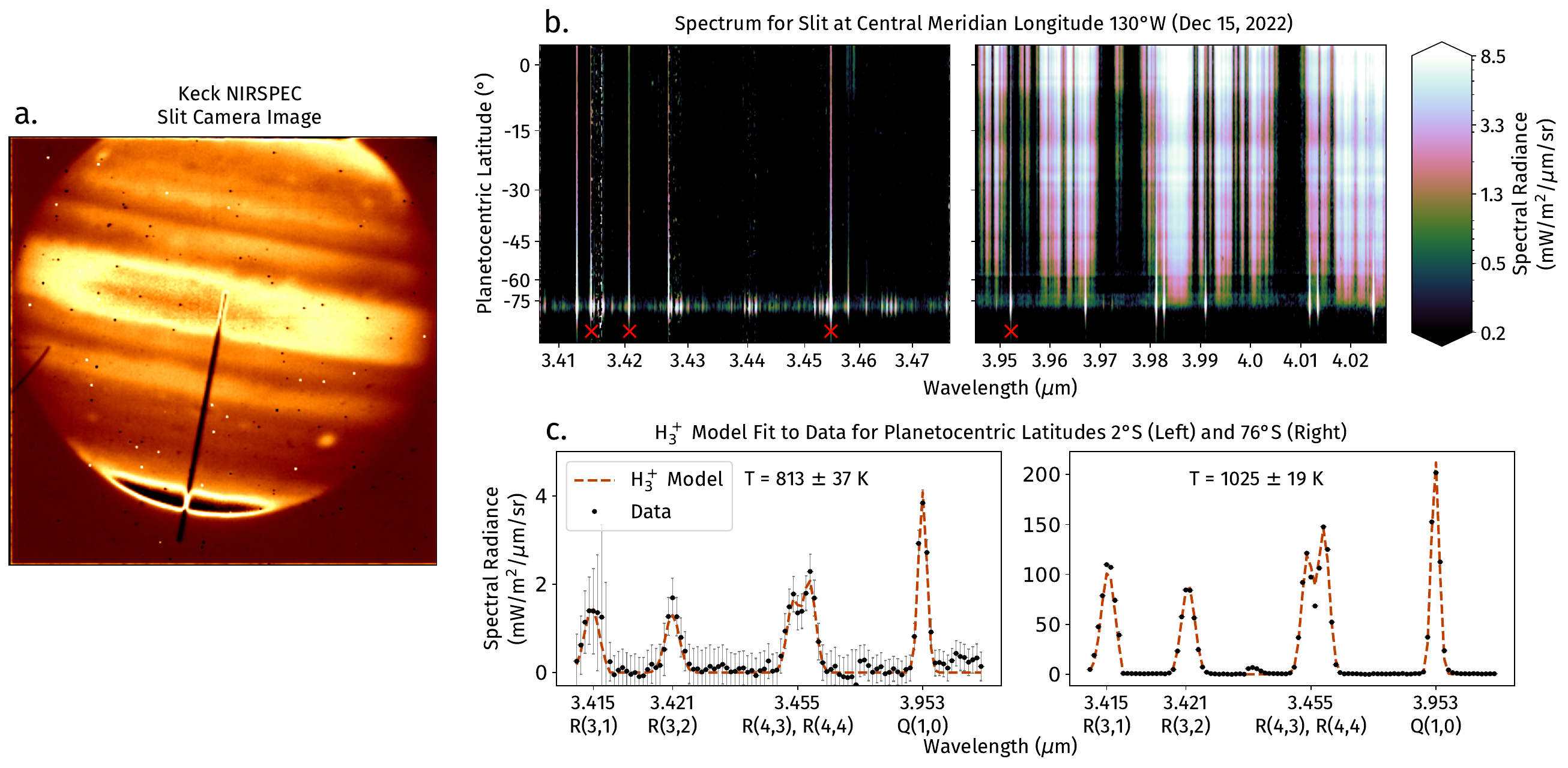}
    \caption{
    \textbf{a.} Jupiter on 15 December 2022 as seen through the NIRSPEC slit camera (SCAM) and shortpass filter (1 - 2.5 $\mu$m). The black slit is aligned along the planet's central meridian, where it passes through the equator and southern aurora before hanging over the southern limb. \textbf{b.} Two orders of a single 2D spectrum corresponding to the SCAM image. We have marked the H$_3^+$ emission lines used in model fitting with red $\times$s. Along the vertical axis, notice the brightening of many H$_3^+$ emissions as they approach the aurora around 75$^{\circ}$ S. There are a number of other H$_3^+$ emissions that were not used in the model fitting, primarily due to telluric and/or methane contamination at non-auroral latitudes. \textbf{c.} Two example \textit{h3ppy} \citep{Melin2020} H$_3^+$ spectral extractions and model fits at the equator (left) and aurora (right) from the same data as the spectrum above. Uncertainties on the data are propagated Poisson noise.
            }
    \label{fig:spectra}
\end{figure}

To effectively map the northern and southern hemispheres of the planet, we aligned the instrument slit perpendicular to the equator along the planet’s Central Meridian Longitude (CML). The 24$''$ slit covered just over one Jupiter radius as seen in in \textbf{Figure \ref{fig:spectra}a} We mapped longitude by nodding from northern to southern slit positions while the planet rotated. Sky spectra were taken every 5 or 6 frames to correct for telluric contamination.

\section{Methods} \label{sec:methods}

\subsection{Data Reduction \label{subsec:red}}

We used the IDL-based reduction package, REDSPEC \citep{Prato2002}, to extract and rectify orders containing viable H$_3^+$ emissions, then performed an initial wavelength calibration using telluric emission lines from sky frames. Non-uniformity across the detector was corrected by flat-fielding. Dark current and Earth’s atmosphere were accounted for with sky subtractions, wherein sky frames taken in regular intervals throughout the observation were matched by closest time to the science spectra of Jupiter. We next performed an absolute flux calibration using A0V stars (HD 1160 for December 2022 and HD 13869 for November/December 2023), which have relatively flat blackbody curves in this wavelength regime, to convert detector counts to physical flux units using the TRDS version of a Kurucz 1993 stellar atmosphere model \citep{LimP2013, Lim2015}. This process included a correction for flux lost due to the narrow width of the spectral slit (0.288$''$), given the point spread function of the star due to seeing (from the MaunaKea Weather Center DIMM instrument: December 2022: 0.57$''$, November 2023: 1.42$''$, December 2023: 1.13$''$). An example of a resulting 2D spectrum from reduction is shown in \textbf{Figure \ref{fig:spectra}b}.

\subsection{Mapping the Spectral Slit in Latitude and Longitude \label{subsec:mapping}}

The three half-nights observed similar ranges of System III West longitude, allowing for comparisons in time of the same region to better judge temporal variability. The CML coverage for each half-night was $\sim$95$^{\circ}$ - 210$^{\circ}$ (December 2022), $\sim$85$^{\circ}$ - 190$^{\circ}$ (November 2023), and $\sim$50$^{\circ}$ - 220$^{\circ}$ (December 2023). This was determined by using NASA Horizons ephemerides to find the CML throughout the time of observation and applying that information in conjunction with SCAM images. The slit geometry on the planet corresponding to each spectrum was determined by using a mapping routine to fit the planetary limb for SCAM images taken concurrently with spectra (\textbf{Figure \ref{fig:spectra}a}). We estimate a typical uncertainty of 0-2 pixels in assigning latitude to spectra along the slit ($\sim$0.5$^{\circ}$ at the equator, $\sim$4$^{\circ}$ near the pole). This follows from a number of possible sources, which include telescope motion during the spectral integration (seen in the 6-8 SCAM frames per spectrum), accuracy in limb fitting, and cropping the spectral order from the wider echelle spectrum. This is similar to uncertainties in previous studies \citep{Kita2018}.

To plot the data, we determine the System III longitude and planetocentric latitude for the center of each pixel along the slit. H$_3^+$ temperatures derived from these spectral frames are then combined into 3$^{\circ}$ by 6$^{\circ}$ latitude-longitude bins to better present and analyze trends. The values in each bin are weighted by temperature uncertainties using a bootstrap median. Assuming a Gaussian distribution of temperatures within each bin, we resample those temperatures and their uncertainties to take a more comprehensive median. Calculated root mean square error propagation uncertainties results in a median error of 2\% for all data, which is representative of the quality of each H$_3^+$ model fit. Temperature variability, calculated from the standard deviation of temperatures within a bin, yields a median variability of 4\% across all data.

\subsection{Temperature Calculations \label{subsec:tempcalc}}

To measure upper atmospheric temperatures we observe infrared emissions from the triatomic hydrogen cation, H$_3^+$, the dominant molecular ion in giant planet ionospheres. When H$_3^+$ is in local thermodynamic equilibrium (LTE) with the surrounding gas its kinetic, rotational, and vibrational temperatures are equal and it acts as an effective proxy of neutral temperature. Early investigation of Jupiter's H$_3^+$ auroral emissions demonstrated that the ion's upper vibrational levels were being populated by thermal processes \citep{Miller1990,Drossart1993}; without spectroscopic information of the H$_3^+$ ground vibrational state it was said to then be in \textit{quasi}-LTE \citep{Miller2020}, in particular for the $\nu_2$ fundamental lines in the 3–4 $\mu$m window.  Subsequent studies indicated that, while full LTE should hold where neutral densities are $\ge$10$^{18}$ m$^{-3}$ \citep{Miller2010}, quasi-LTE should hold up to altitudes of $\sim$800 km above the 1-bar pressure level \citep{Melin2005, Tao2011}. The H$_3^+$ density peak is expected between $\sim$300 and 700 km \citep[e.g.,][]{Nakamura2022,KEDZIORACHUDCZER2017,Egert2017}, (temperature-dependent pressure range: $\sim$1 mbar - 1 $\mu$bar, \cite{Seiff1997}, which is not perfectly representative of all locations on planet). Therefore, H$_3^+$ rotational temperatures derived here are representative of a column-averaged local neutral temperature that is weighted by the H$_3^+$ altitude distribution. For a temperature profile similar to that found by the Galileo Probe \citep{Seiff1997}, observed H$_3^+$ temperatures are modeled to be within 5$\%$ of the neutral temperature at the top of the atmosphere \citep{Moore2019}. 

H$_3^+$ temperatures were calculated using the Python fitting and modeling package, \textit{h3ppy} \citep{Melin2020}. \textit{h3ppy} uses the \cite{Neale1996} H$_3^+$ line list and the \cite{Miller2010} partition function. For this study, we used five H$_3^+$ emission lines that were available across all nights and free from spectral contamination (\textbf{Figure \ref{fig:spectra}c}): 3.41489 $\mu$m R(3,1), 3.42072 $\mu$m R(3,2), a doublet at 3.4547 $\mu$m R(4,3) and 3.45484 $\mu$m R(4,4), and 3.9530 $\mu$m Q(1,0)  \citep{Oka1981}. These emissions appear in the 21$^{st}$ and 24$^{th}$ NIRSPEC orders for our grating settings. The uncertainties on the data in \textbf{Figure \ref{fig:spectra}c} are calculated from their associated Poisson noise - the staunch difference in emission magnitude means error bars are much smaller in the polar regions than equatorial. Other H$_3^+$ emissions are visible in \textbf{Figure \ref{fig:spectra}b}, but were contaminated by telluric and/or methane features across one or more nights. Emission lines are cropped in wavelength from the longer spectrum and a spatial smoothing factor is applied. This smoothing takes the median value of $\pm 3$ pixels in the spatial dimension at a given wavelength to improve H$_3^+$ model fit success. This does not affect the overall median temperatures, but does decrease the standard deviation in a given latitude-longitude bin by $\sim$5 K. We did not apply this smoothing in or near the aurora as the spatial scale of intensity variations is much more abrupt at high latitude.

\begin{figure}
    \centering
    \includegraphics[width=0.9\textwidth]{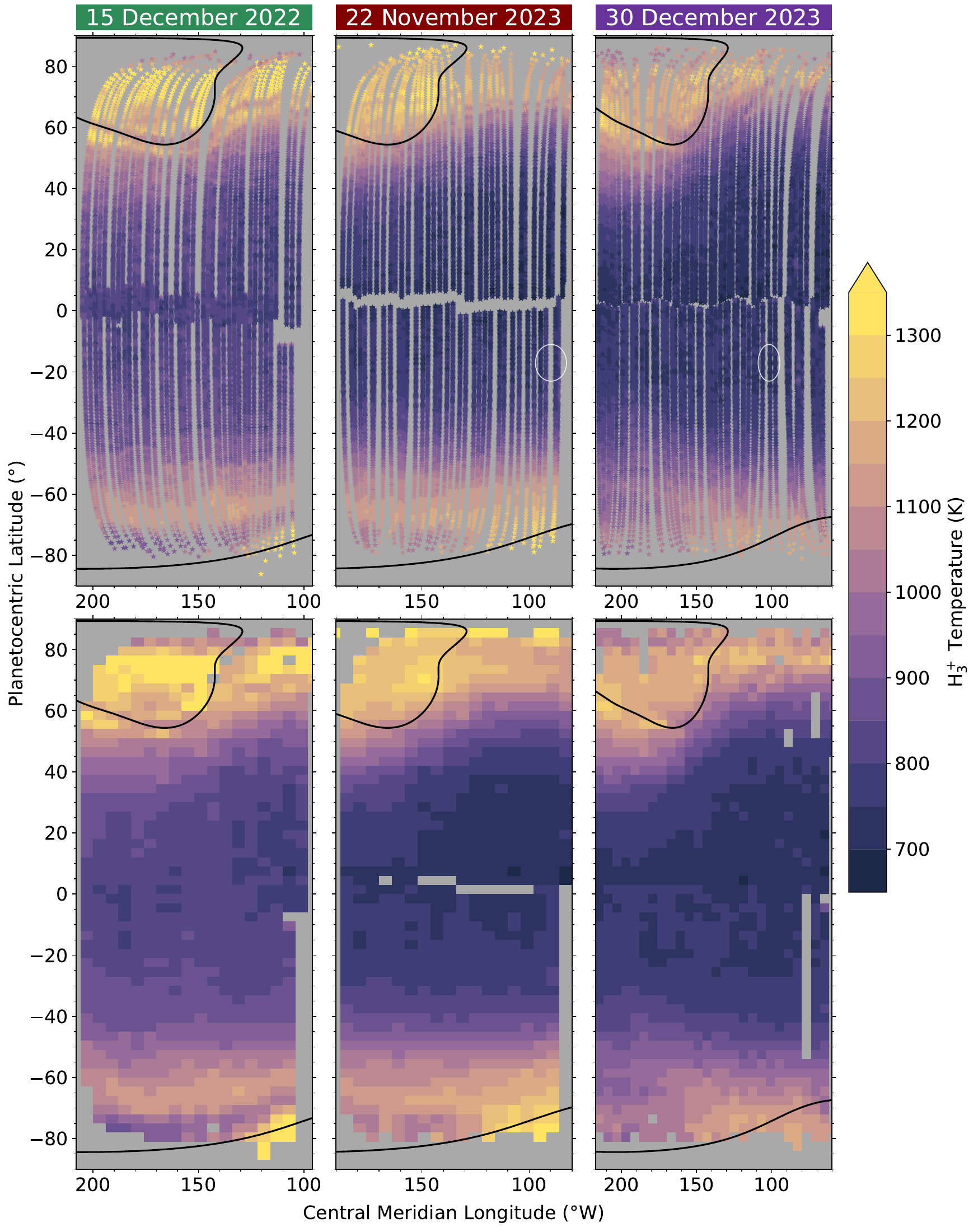}
    \caption{
    Column-averaged H$_3^+$ temperature at Jupiter plotted in planetocentric latitude and central meridian longitude (CML). On each plot, the statistically averaged locations of the outer auroral ovals are plotted \citep{Bonfond2012}. \textbf{Top:} Individual temperature values derived from the center of each pixel along the slit for each night. White ellipses mark the location of the GRS on the November and December 2023 nights. \textbf{Bottom:} The same data binned to 3$^{\circ}$ latitude by 6$^{\circ}$ longitude using a bootstrap median.  
    }
    \label{fig:multimap}
\end{figure}

\section{Results} \label{sec:results}

\textbf{Figure \ref{fig:multimap}} shows mapped temperatures for each night. The top row provides context for the coverage in latitude and longitude, with each point being the calculated center of each pixel along the spectral slit. The bottom row shows the same data binned to 3$^{\circ}$ latitude by 6$^{\circ}$ longitude to more clearly reveal the structure observed in temperature. Each night has been color-coded for figures where the data have been combined; December 2022 is green, November 2023 is red, and December 2023 is purple. The maps also indicate the statistically averaged location of the outer main auroral ovals in solid lines \citep{Bonfond2012}. As indicated with a white ellipse on the upper plots, Jupiter's GRS was observed on November 2023 at a CML of 90$^{\circ}$ and in December 2023 at a CML of 101$^{\circ}$ (both centered at 19$^{\circ}$S latitude); no temperature anomalies are found in those regions, consistent with \cite{Melin2024}. Equatorial temperature, T$_{eq}$, trends are quantified here by calculating the median and standard deviation of all temperature values within $\pm 25^{\circ}$ of the joviographic equator for all observed longitudes on a given night. T$_{eq}$ decreases from $820 \pm 31$ K in December 2022 to $751 \pm 26$ K in November 2023 and $746 \pm 23$ K in December 2023. (T$_{eq}$ differences computed only over longitudes sampled on all three nights are similar.) This could be consistent with a long-term trend of decreasing temperatures, but this dataset is too limited to draw such long-term conclusions.

We also calculate the median and standard deviation of temperature bins strictly within the bounds of the northern (T$_N$) and southern (T$_S$) auroral ovals (i.e. poleward of the solid line at each pole). For December 2022, T$_N$ = $1247 \pm 122$ K and T$_S$ = $1357 \pm 128$ K. For November 2023, T$_N$ = $1230 \pm 55$ K and T$_S$ = $1295 \pm 18$ K. For December 2023, T$_N$ = $1163 \pm 56$ K and T$_S$ = $1130 \pm 62$ K. Due to the southern oval being well-centered at the planetographic pole and Jupiter's small obliquity, we have limited coverage of the southern aurora and thus these values have been calculated from a small number of temperatures, especially December 2022 which only has ten observed temperature values. Comparing values from all three nights, we observe a median northern temperature of $1200 \pm 92$ K from 1727 auroral values and a southern temperature of $1143 \pm 120$ K from 96 auroral values. Therefore, there is no clear distinction between northern and southern auroral temperatures, at least within this limited dataset.

\begin{figure}
    \centering
    \includegraphics[width=0.99\textwidth]{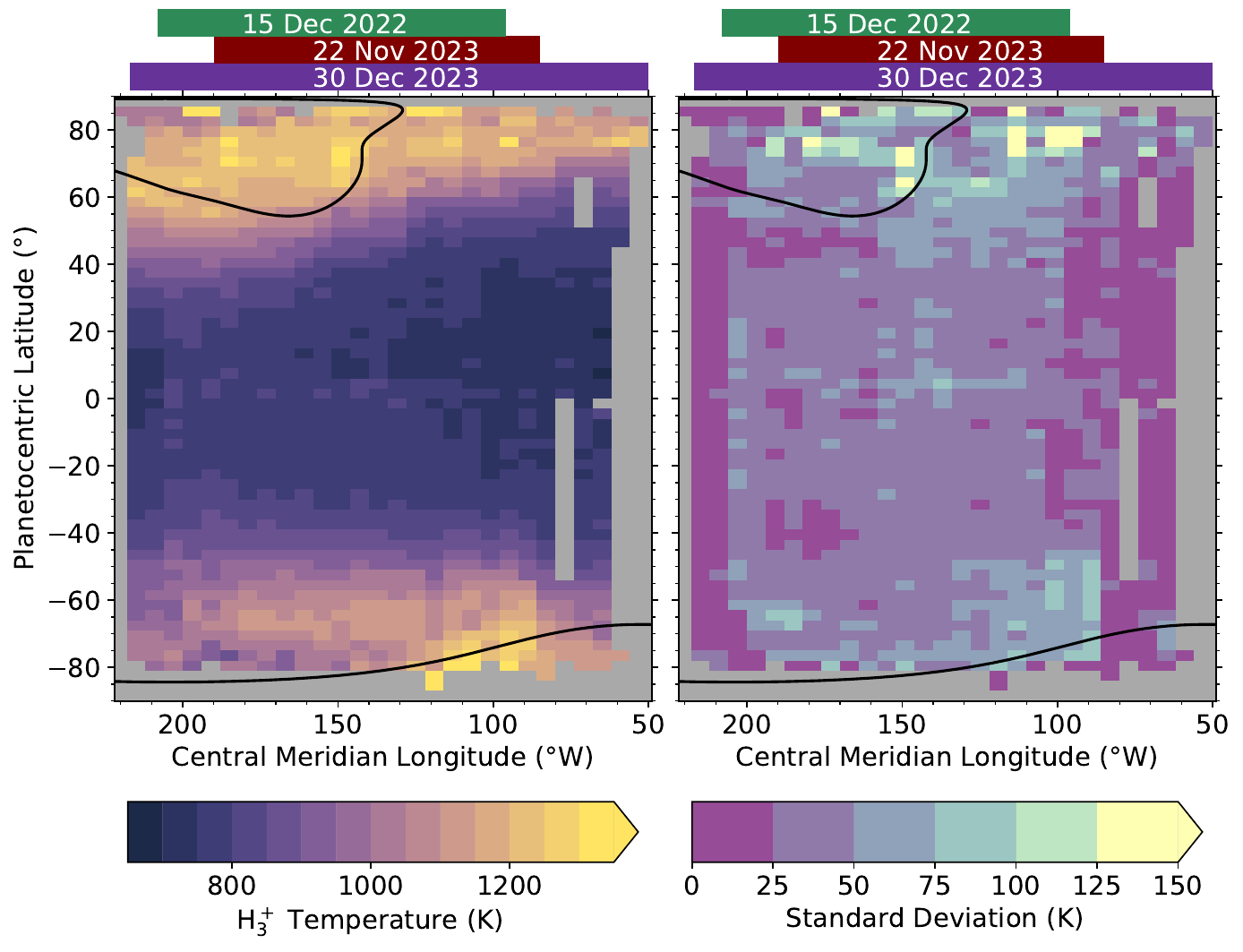}
    \caption{
    Column-averaged H$_3^+$ temperature and corresponding standard deviation. Data have been binned to 3$^{\circ}$ latitude by 6$^{\circ}$ longitude. Statistically averaged locations of the outer main auroral ovals are shown in solid lines \citep{Bonfond2012}, with the longitudinal coverage of each night of observation indicated at the top in green (December 2022), red (November 2023), and purple (December 2023).
    }
    \label{fig:combined}
\end{figure}

In \textbf{Figure \ref{fig:combined}}, data from all three nights have been combined into 3$^{\circ}$ by 6$^{\circ}$ latitude-longitude bins. The final temperature of a bin is the bootstrap median of all values within the bin which accounts for the associated uncertainties from the model fits. The H$_3^+$ emission line model fits used to calculate these temperatures have a median root mean square error of 2\% and a maximum of 11\%, with higher errors at low-latitude, as indicated in \textbf{Figure \ref{fig:spectra}c}. The corresponding standard deviation of the binned temperatures is mapped on the right: the median deviation is 4\% and the maximum is 21\%. Along the top of each map are colored bars which denote the longitudinal coverage of each night; following their assigned colors, December 2022 is green, November 2023 is red, and December 2023 is purple. As before, the statistically averaged location of the outer main auroral ovals from \cite{Bonfond2012} are shown with solid lines.

\begin{figure}
    \centering
    \includegraphics[width=0.9\textwidth]{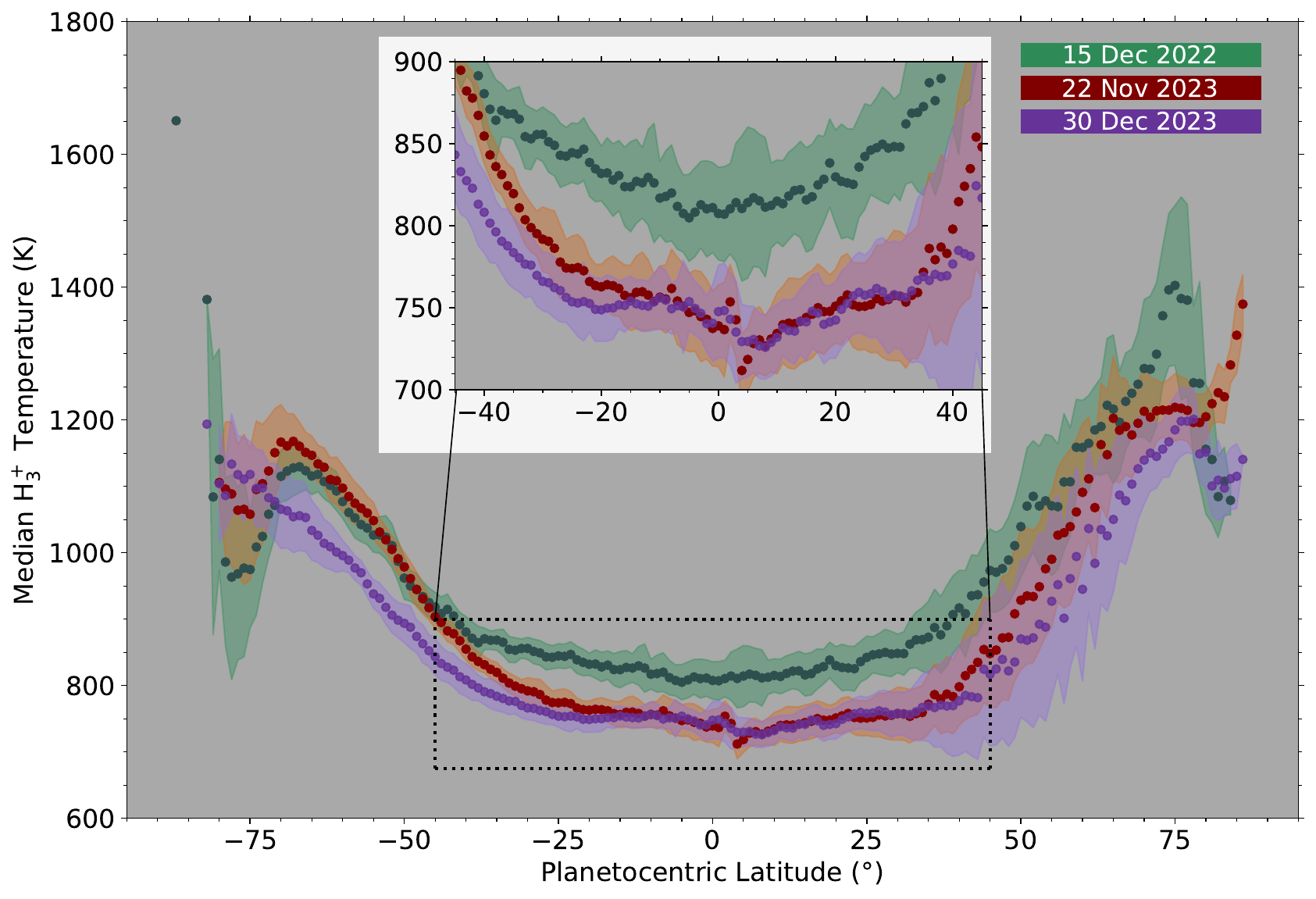}
    \caption{
    H$_3^+$ column-averaged temperature profiles for all three nights (December 2022 in green, November 2023 in red, and December 2023 in purple). These profiles represent the median temperature, binned by 1$^{\circ}$ in latitude, across all observed longitudes. The inset plot improves visibility for the similarities and differences between the three nights at equatorial latitudes. Shaded regions show the standard deviation in each latitude bin. 
    }
    \label{fig:summed}
\end{figure}

In \textbf{Figure \ref{fig:summed}}, data have been collapsed in longitude to observe latitudinal trends in temperature from night-to-night. Each night is plotted in its corresponding colors. The standard deviation of all temperatures within each 1$^{\circ}$ latitude bin is represented in the shaded boundary. From this, the differences between observations becomes more apparent: December 2022 is significantly hotter in the equatorial region, while the two nights in 2023 agree relatively well. We note the presence of a transient sub-auroral temperature minimum around 75$^{\circ}$ S in December 2022 and November 2023 which is also present but less obvious in \textbf{Figure \ref{fig:multimap}}. In the online version of this paper, we have included an animated version of \textbf{Figure \ref{fig:summed}} which shows how latitudinal temperature structure changes over 6$^{\circ}$ longitude bins for each night of observation.

\section{Discussion} \label{sec:disc}

\subsection{Auroral Energetics \& the Role of H$_3^+$ \label{subsec:auroraeng}}

Observed temperatures here lie within the wide range of previously published  values. For the equatorial region, ground-based literature includes typical column-averaged H$_3^+$ temperatures from 500 K \citep{Stallard2017} to 800 K \citep{ODonoghue2016}, while Juno/JIRAM has observed temperatures of $\sim$800 K \citep{Migliorini2019}. In the aurora, ground-based observations of northern temperatures range from 650 K \citep{Miller1997, Ballester1994} to 1200 K \citep{Raynaud2004}, with spacecraft observations spanning 500 K \citep[Galileo/NIMS;][]{Altieri2016} to 1400 K \citep[Cassini/VIMS;][]{Stallard2015}. For the southern aurora, ground-based observations range from 850 K \citep{Kita2018} to 1250 K \citep{Lystrup2008} and spacecraft have observed 940 K  \citep[Juno/JIRAM;][]{Adriani2017} to 1200 K \citep[Cassini/VIMS;][]{Stallard2015}. 

The substantial variation of reported auroral temperatures, and the relatively high auroral temperature variability found here, is reasonable when considering the stochastic nature of those emissions \citep[e.g.,][]{Badman2015}. In addition to spatial and temporal variations of auroral particle precipitation, column-averaged H$_3^+$ temperatures are also sensitive to the mean energy of any precipitating particles. Whereas the primary difference in ultraviolet (UV) emission with increasing mean energy is that there will be more absorption due to methane and a higher derived color ratio \citep{Gerard2023}, H$_3^+$ emission is weighted by its vertical density distribution. For example, precipitating electrons with mean energy of 100 keV would produce enhanced ionization peaking near 300 km altitude \citep{Tao2011,Gerard2014}, meaning column-averaged H$_3^+$ temperatures in that region would primarily reflect temperatures near the base of the thermosphere. In contrast, a high latitude solar-produced ionosphere would generate a relatively high altitude H$_3^+$ density peak due to the increased atmosphere path length. Derived temperatures would then be higher for a typical Bates-like profile with a positive thermospheric temperature gradient \citep[e.g.,][]{Seiff1997}. Alternatively, where auroral heating mechanisms generate low altitude ($\sim$400 km, $\sim$1 $\mu$bar) hot spots as predicted by models \citep{Majeed2009,Yates2020}, then higher energy precipitation would weight observed H$_3^+$ emission to the low-altitude, high-temperature regions. In either case, while H$_3^+$ upper atmospheric temperatures are observed to be highest on average in the auroral region, their high degree of spatiotemporal variability helps to emphasize that, on their own, they cannot be used to reliably constrain auroral energetics. Additional information, such as altitude profiles of H$_3^+$ temperature and volumetric number density, is needed to contextualize observed variability.

\subsection{Determining a Dominant, Upper-Atmospheric Jovian Energy Source \label{subsec:energysource}}

Global temperature trends are more revealing of the dominant energy source(s) in Jupiter's upper atmosphere. If heating due to upward-propagating, lower-atmospheric waves is significant, there should be localized temperature hot spots above any active regions. We see no such evidence across three half-nights of observation, indicating that wave heating is not contributing significantly to the upper atmospheric energy budget over this period or that perhaps it is a distributed low-level amount of heating. Instead, temperatures peak at high latitude, especially within the statistically averaged auroral oval locations, and decrease steadily towards the equator. This trend is expected were Jupiter's aurorae the dominant source of upper atmospheric energy \citep[e.g.,][]{Bougher2005,Tao2009}. Such a temperature gradient has been seen before at Jupiter \citep{ODonoghue2021}. By demonstrating a repeating temperature gradient over the same range of longitudes across three separate nights, these results significantly expand on the observational evidence in support of the notion that Jupiter's non-auroral latitudes are heated primarily by dynamical redistribution of auroral energy. 

There are, however, a few exceptions to such a latitudinal trend in the literature. Pole-to-pole H$_3^+$ temperature measurements by \cite{Ballester1994} found low-latitude temperature enhancements up to 1200 K near 102$^{\circ}$ W CML. \cite{Lam1997} produced (interpolated) global maps of observed H$_3^+$ parameters, and found similar low-latitude hot spots, of order 900 K, at a range of CMLs. More recently, \cite{Migliorini2019} derived H$_3^+$ altitude profiles from Juno/JIRAM limb observations and found that temperatures were generally coolest at mid-latitude and increased towards the equator, while \cite{ODonoghue2016} found a $\sim$1600 K hot spot above the GRS. 

Taken together, it seems that direct upper atmospheric heating by upward propagating waves is, at best, variable. It may lead to localized hot spots, but they appear to be either short-lived, infrequent, or both. In particular, the absence of any non-auroral hot spots on either 22 November or 30 December 2023 (\textbf{Figure \ref{fig:multimap}}) implies that either direct wave heating signatures persist for $<$38 days, no such “events" occurred over that period in the observed longitude sector, or they are occurring on smaller spatial or temperature scales than we can resolve. Recent JWST/NIRSpec observations revealed evidence of such small scale wave structure surrounding the GRS \citep{Melin2024}. In the meantime, observed temperature gradients demonstrate that the Coriolis barrier responsible for trapping auroral energy at high latitudes in some GCMs must somehow be overcome by Jupiter. A similar temperature trend was derived from UV occultations at Saturn \citep{Brown2020}, where zonal drag due to for atmospheric gravity waves -- seen both in remote temperature profiles and in situ Cassini measurements -- was used to enable equatorward redistribution of auroral energy in a Saturn GCM \citep{Mueller-Wodarg2019}. Therefore, while it is clear that waves are present in giant planet thermospheres, their primary contribution to upper atmospheric heating may be in modifying dynamics. In that scenario, prior observations of infrequent non-auroral H$_3^+$ temperature hot spots might be evidence that, while upward propagating waves may impart ephemeral heating, they do not contribute significantly to the long-term energy budget of Jupiter's low-latitude upper atmosphere.

\subsection{A Longitudinal Anomaly \label{subsec:oddity}}

Beyond just a monotonic decrease in H$_3^+$ temperature from polar to equatorial latitudes, these maps reveal a possible organization of temperature with System III West longitude, and thus, magnetic field. For these observations the spectral slit was centered along the central meridian, so they are all fixed near local solar noon. Yet, even on a single night of observing, individual noon equatorial ($\pm$25$^{\circ}$) temperature measurements are observed to vary by more than 200 K as a function of CML (as seen in the top panels of \textbf{Figure \ref{fig:multimap}}). In simulations, neutral winds diverging from the auroral region are rapidly turned westward by strong Coriolis forces, meaning the mid- to low-latitude thermosphere is expected to exhibit strong zonal winds (decreasing with latitude) with minimal variation in longitude/local time. Symmetry created in longitude by zonal winds is strong enough that GCMs present temperatures as zonally averaged, i.e. no temperature structures corresponding to those in \textbf{Figures \ref{fig:multimap}} and \textbf{\ref{fig:combined}} are found even for model results presented in latitude and longitude \citep[e.g.,][]{Majeed2009} (though it must be noted that Jupiter GCMs have not yet incorporated JRM33, the Juno-derived magnetic field model that includes finer-scale magnetic field features, e.g. \cite{Bougher2005}, \cite{Tao2009}, \cite{Yates2020}). Thus, based on existing simulations, the observed longitudinal temperature variations appear to be unlikely to represent true variations in thermospheric temperature. Due to the nature of H$_3^+$ temperature retrievals, which are weighted near the H$_3^+$ density peak in altitude, they might instead be associated with a low-altitude ionospheric region.

To create a region of low-altitude H$_3^+$ density, we now turn to local electrodynamics and Jupiter’s magnetic field. Around 90$^{\circ}$ W, Jupiter’s magnetic field is very complex, possibly related to a magnetic oddity known as the “Great Blue Spot" \citep{Moore2018, Connerney2022}. Associated plasma dynamics are also more complicated within this region of twisted magnetic field. In a perfect dipole, magnetic field lines are oriented perpendicular to the flow of neutral zonal winds. In this region, however, field lines have a more significant zonal component which could cause westward neutral winds to drive plasma down field lines, and thus down in altitude, at northern latitudes. This would be consistent with a preferential north-south asymmetry in temperature near 90$^{\circ}$ W. 

The region of H$_3^+$ temperature decrease also coincides with H$_3^+$ emission features found using images and spectra taken more than 20 years ago \citep{Stallard2018, Drossart2019} as well as a localized “bulge" of enhanced H Ly-$\alpha$ emission discovered more than 40 years ago \citep{Clarke1980,Sandel1980,Melin2016}. \cite{Stallard2018} presented a global map constructed from thousands of narrow-band images which found evidence for localized interactions with Jupiter's magnetic field. Two prominent features stand out from that map: a sinusoidal pattern of weakened H$_3^+$ emission (known as the “Dark Ribbon" and shown to track the magnetic equator), and a large region of dimmed H$_3^+$ emission between $\sim$60 and 150$^{\circ}$ W CML which is coincident with the “Great Blue Spot”. H$_3^+$ being pushed to lower altitudes is an attractive explanation for our observed temperatures as it may also relate to these broader mysteries. Rather than recreating a broad H Ly-$\alpha$ line profile in the bulge region \citep{Clarke1980} by dissociative recombination of H$_3^+$ with precipitating electrons \citep{Melin2016}, a forced H$_3^+$ downwelling could cause H$_3^+$ to charge-exchange with methane. This would lead to a chain of hydrocarbon ions, some of which may also produce hot H upon recombination, leading to the broadened line profile observed by \cite{Clarke1980} and eliminating the need for low-latitude electron precipitation presented in \cite{Melin2016}. Recombination would also result in a local depletion in electrons, which is present in Galileo radio occultations and Juno \textit{in situ} data in this region \citep{Mendillo2022, Kurth2025}. Given the correspondence of depleted H$_3^+$ emission, the H Ly-$\alpha$ bulge, the unusual magnetic field structure, and now our temperature measurements  at 20$^{\circ}$ N, 90$^{\circ}$ over varying time scales at 20$^{\circ}$ N, 90$^{\circ}$, it seems that these oddities are likely not associated with a variable external driver, such as solar EUV flux or possible low-latitude particle precipitation.

\section{Conclusions} \label{sec:conclusions}

We present the first multi-epoch, pole-to-pole  high resolution H$_3^+$ temperature map as a first step in constraining the temperatures in Jupiter's upper atmosphere. We find a median equatorial ($\pm 25^{\circ}$) temperature of T$_{eq}$ = $762 \pm 43$ K over three half-nights of observation covering similar magnetic longitudes separated by time spans of one month and one year. While median equatorial temperatures can differ by $\sim$75 K in one year, they are remarkably consistent otherwise. Median equatorial temperatures vary only by 5 K in just over a month and the median non-auroral temperature deviation is only 4$\%$ across $\sim$40 data points in each equatorial latitude-longitude bin. 

Observed H$_3^+$ temperatures decrease from aurora to equator, and we see no evidence for any localized non-auroral hot spots. These smooth global temperature gradients are consistent with the aurorae being the primary source of heating for Jupiter's non-auroral upper atmosphere. 
In addition, we find that mid- and low-latitude H$_3^+$ temperatures surrounding the 20$^{\circ}$ N, 90$^{\circ}$ W region are systematically cooler ($\sim$25 K) than those near 180$^{\circ}$ W. This apparent magnetic control of H$_3^+$ temperature (and emission, as detected previously) is coterminous with other unusual features at Jupiter: notably the complex magnetic field surrounding the “Great Blue Spot" and the H Ly-$\alpha$ bulge. 

Future work will focus on expanding this map by coverage in time and space. With a fully global map, we can develop significantly improved constraints on the energetics of Jupiter's upper atmosphere as a whole rather than over a subset of longitudes. By expanding our temporal coverage, we can better understand Jovian temperature variability and its possible drivers.

\section{Acknowledgments}

KR was supported by NASA FINESST Grant 80NSSC23K1637. Observations were made possible with Grant 80NSSC22K095 from NASA's Solar System Observations program. HM was supported by the STFC James Webb Fellowship (ST/W001527/2) at Northumbria University, UK. KLK was supported by a Northumbria University Research Studentship at Northumbria University, UK. PIT was supported by UK Science and Technology Facilities Council (STFC) Studentship ST/X508548/2. 

This research was selected under the Key Strategic Mission Support (KSMS) category of NASA’s Keck General Observing program which is managed by the NASA Exoplanet Science Institute (NExScI) and funded under the auspices of NASA’s Planetary Science Division. Data presented herein were obtained at the W. M. Keck Observatory from telescope time allocated to the National Aeronautics and Space Administration through the agency's scientific partnership with the California Institute of Technology and the University of California. The Observatory was made possible by the generous financial support of the W. M. Keck Foundation. The authors wish to recognize and acknowledge the very significant cultural role and reverence that the summit of Mauna Kea has always had within the indigenous Hawaiian community. We are most fortunate to have the opportunity to conduct observations from this mountain.

This research was supported by the International Space Science Institute (ISSI) for the international team on “Jupiter’s Non-Auroral Ionosphere”.

\vspace{5mm}
\facility{Keck:II (NIRSPEC)}

\software{h3ppy
\citep{Melin2020},
          Astropy 
\citep{2013A&A...558A..33A,2018AJ....156..123A},
          pandas
\citep{mckinney-proc-scipy-2010,reback2020pandas},
          SciPy
\citep{2020SciPy-NMeth},
          PySynphot
\citep{Lim2015, LimP2013},
          numpy
\citep{Harris2020},
          Matplotlib
\citep{Hunter:2007}
          }




\bibliography{Paper1bib}{}

\begin{thebibliography}{}
\expandafter\ifx\csname natexlab\endcsname\relax\def\natexlab#1{#1}\fi
\providecommand{\url}[1]{\href{#1}{#1}}
\providecommand{\dodoi}[1]{doi:~\href{http://doi.org/#1}{\nolinkurl{#1}}}
\providecommand{\doeprint}[1]{\href{http://ascl.net/#1}{\nolinkurl{http://ascl.net/#1}}}
\providecommand{\doarXiv}[1]{\href{https://arxiv.org/abs/#1}{\nolinkurl{https://arxiv.org/abs/#1}}}

\bibitem[{Adriani {et~al.}(2017)Adriani, Mura, Moriconi, Dinelli, Fabiano, Altieri, Sindoni, Bolton, Connerney, Atreya, Bagenal, Gérard, Filacchione, Tosi, Migliorini, Grassi, Piccioni, Noschese, Cicchetti, Gladstone, Hansen, Kurth, Levin, Mauk, McComas, Olivieri, Turrini, Stefani, \& Amoroso}]{Adriani2017}
Adriani, A., Mura, A., Moriconi, M.~L., {et~al.} 2017, Geophysical Research Letters, 44, 4633, \dodoi{10.1002/2017GL072905}

\bibitem[{Altieri {et~al.}(2016)Altieri, Dinelli, Migliorini, Moriconi, Sindoni, Adriani, Mura, \& Fabiano}]{Altieri2016}
Altieri, F., Dinelli, B.~M., Migliorini, A., {et~al.} 2016, Geophysical Research Letters, 43, 11,558, \dodoi{10.1002/2016GL070787}

\bibitem[{{Astropy Collaboration} {et~al.}(2013){Astropy Collaboration}, {Robitaille}, {Tollerud}, {Greenfield}, {Droettboom}, {Bray}, {Aldcroft}, {Davis}, {Ginsburg}, {Price-Whelan}, {Kerzendorf}, {Conley}, {Crighton}, {Barbary}, {Muna}, {Ferguson}, {Grollier}, {Parikh}, {Nair}, {Unther}, {Deil}, {Woillez}, {Conseil}, {Kramer}, {Turner}, {Singer}, {Fox}, {Weaver}, {Zabalza}, {Edwards}, {Azalee Bostroem}, {Burke}, {Casey}, {Crawford}, {Dencheva}, {Ely}, {Jenness}, {Labrie}, {Lim}, {Pierfederici}, {Pontzen}, {Ptak}, {Refsdal}, {Servillat}, \& {Streicher}}]{2013A&A...558A..33A}
{Astropy Collaboration}, {Robitaille}, T.~P., {Tollerud}, E.~J., {et~al.} 2013, \aap, 558, A33, \dodoi{10.1051/0004-6361/201322068}

\bibitem[{{Astropy Collaboration} {et~al.}(2018){Astropy Collaboration}, {Price-Whelan}, {Sip{\H{o}}cz}, {G{\"u}nther}, {Lim}, {Crawford}, {Conseil}, {Shupe}, {Craig}, {Dencheva}, {Ginsburg}, {VanderPlas}, {Bradley}, {P{\'e}rez-Su{\'a}rez}, {de Val-Borro}, {Aldcroft}, {Cruz}, {Robitaille}, {Tollerud}, {Ardelean}, {Babej}, {Bach}, {Bachetti}, {Bakanov}, {Bamford}, {Barentsen}, {Barmby}, {Baumbach}, {Berry}, {Biscani}, {Boquien}, {Bostroem}, {Bouma}, {Brammer}, {Bray}, {Breytenbach}, {Buddelmeijer}, {Burke}, {Calderone}, {Cano Rodr{\'\i}guez}, {Cara}, {Cardoso}, {Cheedella}, {Copin}, {Corrales}, {Crichton}, {D'Avella}, {Deil}, {Depagne}, {Dietrich}, {Donath}, {Droettboom}, {Earl}, {Erben}, {Fabbro}, {Ferreira}, {Finethy}, {Fox}, {Garrison}, {Gibbons}, {Goldstein}, {Gommers}, {Greco}, {Greenfield}, {Groener}, {Grollier}, {Hagen}, {Hirst}, {Homeier}, {Horton}, {Hosseinzadeh}, {Hu}, {Hunkeler}, {Ivezi{\'c}}, {Jain}, {Jenness}, {Kanarek}, {Kendrew}, {Kern}, {Kerzendorf}, {Khvalko}, {King}, {Kirkby}, {Kulkarni},
  {Kumar}, {Lee}, {Lenz}, {Littlefair}, {Ma}, {Macleod}, {Mastropietro}, {McCully}, {Montagnac}, {Morris}, {Mueller}, {Mumford}, {Muna}, {Murphy}, {Nelson}, {Nguyen}, {Ninan}, {N{\"o}the}, {Ogaz}, {Oh}, {Parejko}, {Parley}, {Pascual}, {Patil}, {Patil}, {Plunkett}, {Prochaska}, {Rastogi}, {Reddy Janga}, {Sabater}, {Sakurikar}, {Seifert}, {Sherbert}, {Sherwood-Taylor}, {Shih}, {Sick}, {Silbiger}, {Singanamalla}, {Singer}, {Sladen}, {Sooley}, {Sornarajah}, {Streicher}, {Teuben}, {Thomas}, {Tremblay}, {Turner}, {Terr{\'o}n}, {van Kerkwijk}, {de la Vega}, {Watkins}, {Weaver}, {Whitmore}, {Woillez}, {Zabalza}, \& {Astropy Contributors}}]{2018AJ....156..123A}
{Astropy Collaboration}, {Price-Whelan}, A.~M., {Sip{\H{o}}cz}, B.~M., {et~al.} 2018, \aj, 156, 123, \dodoi{10.3847/1538-3881/aabc4f}

\bibitem[{{Badman} {et~al.}(2015){Badman}, {Branduardi-Raymont}, {Galand}, {Hess}, {Krupp}, {Lamy}, {Melin}, \& {Tao}}]{Badman2015}
{Badman}, S.~V., {Branduardi-Raymont}, G., {Galand}, M., {et~al.} 2015, \ssr, 187, 99, \dodoi{10.1007/s11214-014-0042-x}

\bibitem[{Ballester {et~al.}(1994)Ballester, Miller, Tennyson, Trafton, \& Geballe}]{Ballester1994}
Ballester, G.~E., Miller, S., Tennyson, J., Trafton, L.~M., \& Geballe, T.~R. 1994, Icarus, 107, 189, \dodoi{10.1006/icar.1994.1015}

\bibitem[{Bonfond {et~al.}(2012)Bonfond, Grodent, Gérard, Stallard, Clarke, Yoneda, Radioti, \& Gustin}]{Bonfond2012}
Bonfond, B., Grodent, D., Gérard, J.~C., {et~al.} 2012, Geophysical Research Letters, 39, \dodoi{10.1029/2011GL050253}

\bibitem[{Bougher {et~al.}(2005)Bougher, Waite, Majeed, \& Gladstone}]{Bougher2005}
Bougher, S.~W., Waite, J.~H., Majeed, T., \& Gladstone, G.~R. 2005, Journal of Geophysical Research: Planets, 110, 1, \dodoi{10.1029/2003JE002230}

\bibitem[{{Brown} {et~al.}(2020){Brown}, {Koskinen}, {M{\"u}ller-Wodarg}, {West}, {Jouchoux}, \& {Esposito}}]{Brown2020}
{Brown}, Z., {Koskinen}, T., {M{\"u}ller-Wodarg}, I., {et~al.} 2020, Nature Astronomy, 4, 872, \dodoi{10.1038/s41550-020-1060-0}

\bibitem[{{Clarke} {et~al.}(1980){Clarke}, {Weaver}, {Feldman}, {Moos}, {Fastie}, \& {Opal}}]{Clarke1980}
{Clarke}, J.~T., {Weaver}, H.~A., {Feldman}, P.~D., {et~al.} 1980, \apj, 240, 696, \dodoi{10.1086/158277}

\bibitem[{Connerney {et~al.}(2022)Connerney, Timmins, Oliversen, Espley, Joergensen, Kotsiaros, Joergensen, Merayo, Herceg, Bloxham, Moore, Mura, Moirano, Bolton, \& Levin}]{Connerney2022}
Connerney, J.~E., Timmins, S., Oliversen, R.~J., {et~al.} 2022, Journal of Geophysical Research: Planets, 127, \dodoi{10.1029/2021JE007055}

\bibitem[{Drossart(2019)}]{Drossart2019}
Drossart, P. 2019, Philosophical Transactions of the Royal Society A, 377, \dodoi{10.1098/RSTA.2018.0404}

\bibitem[{{Drossart} {et~al.}(1993){Drossart}, {Maillard}, {Caldwell}, \& {Rosenqvist}}]{Drossart1993}
{Drossart}, P., {Maillard}, J.~P., {Caldwell}, J., \& {Rosenqvist}, J. 1993, \apjl, 402, L25, \dodoi{10.1086/186691}

\bibitem[{{Egert} {et~al.}(2017){Egert}, {Waite}, \& {Bell}}]{Egert2017}
{Egert}, A., {Waite}, J.~H., \& {Bell}, J. 2017, Journal of Geophysical Research (Space Physics), 122, 2210, \dodoi{10.1002/2016JA023189}

\bibitem[{{G{\'e}rard} {et~al.}(2014){G{\'e}rard}, {Bonfond}, {Grodent}, {Radioti}, {Clarke}, {Gladstone}, {Waite}, {Bisikalo}, \& {Shematovich}}]{Gerard2014}
{G{\'e}rard}, J.~C., {Bonfond}, B., {Grodent}, D., {et~al.} 2014, Journal of Geophysical Research (Space Physics), 119, 9072, \dodoi{10.1002/2014JA020514}

\bibitem[{{G{\'e}rard} {et~al.}(2023){G{\'e}rard}, {Gkouvelis}, {Bonfond}, {Gladstone}, {Mura}, {Adriani}, {Grodent}, {Hue}, \& {Greathouse}}]{Gerard2023}
{G{\'e}rard}, J.~C., {Gkouvelis}, L., {Bonfond}, B., {et~al.} 2023, \icarus, 389, 115261, \dodoi{10.1016/j.icarus.2022.115261}

\bibitem[{Harris {et~al.}(2020)Harris, Millman, van~der Walt, Gommers, Virtanen, Cournapeau, Wieser, Taylor, Berg, Smith, Kern, Picus, Hoyer, van Kerkwijk, Brett, Haldane, del Río, Wiebe, Peterson, Gérard-Marchant, Sheppard, Reddy, Weckesser, Abbasi, Gohlke, \& Oliphant}]{Harris2020}
Harris, C.~R., Millman, K.~J., van~der Walt, S.~J., {et~al.} 2020, Nature 2020 585:7825, 585, 357, \dodoi{10.1038/s41586-020-2649-2}

\bibitem[{{Hubbard} {et~al.}(1972){Hubbard}, {Nather}, {Evans}, {Tull}, {Wells}, {van Citters}, {Warner}, \& {vanden Bout}}]{Hubbard1972}
{Hubbard}, W., {Nather}, R.~E., {Evans}, D.~S., {et~al.} 1972, \aj, 77, 41, \dodoi{10.1086/111244}

\bibitem[{Hunter(2007)}]{Hunter:2007}
Hunter, J.~D. 2007, Computing in Science \& Engineering, 9, 90, \dodoi{10.1109/MCSE.2007.55}

\bibitem[{Kedziora-Chudczer {et~al.}(2017)Kedziora-Chudczer, Cotton, Kedziora, \& Bailey}]{KEDZIORACHUDCZER2017}
Kedziora-Chudczer, L., Cotton, D., Kedziora, D., \& Bailey, J. 2017, Icarus, 294, 156, \dodoi{https://doi.org/10.1016/j.icarus.2017.04.029}

\bibitem[{Kita {et~al.}(2018)Kita, Fujisawa, Tao, Kagitani, Sakanoi, \& Kasaba}]{Kita2018}
Kita, H., Fujisawa, S., Tao, C., {et~al.} 2018, Icarus, 313, 93, \dodoi{10.1016/J.ICARUS.2018.05.002}

\bibitem[{Kurth {et~al.}(2025)Kurth, Faden, Sulaiman, Elliott, Hospodarsky, Connerney, Kammer, Greathouse, Valek, Allegrini, Bagenal, Stallard, Moore, Coffin, Agiwal, Withers, \& Bolton}]{Kurth2025}
Kurth, W., Faden, J., Sulaiman, A., {et~al.} 2025, JGR: Planets

\bibitem[{Lam {et~al.}(1997)Lam, Achilleos, Miller, Tennyson, Trafton, Geballe, \& Ballester}]{Lam1997}
Lam, H.~A., Achilleos, N., Miller, S., {et~al.} 1997, Icarus, 127, 379, \dodoi{10.1006/ICAR.1997.5698}

\bibitem[{{Lian} \& {Yelle}(2019)}]{Lian2019}
{Lian}, Y., \& {Yelle}, R.~V. 2019, \icarus, 329, 222, \dodoi{10.1016/j.icarus.2019.04.001}

\bibitem[{Lim {et~al.}(2015)Lim, Diaz, \& Laidler}]{Lim2015}
Lim, P., Diaz, R., \& Laidler, V. 2015, pysynphot: Synthetic photometry software package.
\newblock \url{http://ascl.net/1303.023}

\bibitem[{Lim {et~al.}(2013)Lim, Diaz, \& Laidler}]{LimP2013}
Lim, P.~L., Diaz, R.~I., \& Laidler. 2013, pysynphot: Synthetic photometry software package.
\newblock \url{https://ui.adsabs.harvard.edu/abs/2013ascl.soft03023S/abstract}

\bibitem[{Lystrup {et~al.}(2008)Lystrup, Miller, Russo, R.~J.~Vervack, \& Stallard}]{Lystrup2008}
Lystrup, M.~B., Miller, S., Russo, N.~D., R.~J.~Vervack, J., \& Stallard, T. 2008, The Astrophysical Journal, 677, 790, \dodoi{10.1086/529509/FULLTEXT/}

\bibitem[{{Majeed} {et~al.}(2009){Majeed}, {Waite}, {Bougher}, \& {Gladstone}}]{Majeed2009}
{Majeed}, T., {Waite}, J.~H., {Bougher}, S.~W., \& {Gladstone}, G.~R. 2009, Journal of Geophysical Research (Planets), 114, E07005, \dodoi{10.1029/2008JE003194}

\bibitem[{Martin {et~al.}(2016)Martin, Fitzgerald, McLean, Kress, \& Wang}]{Martin2016}
Martin, E.~C., Fitzgerald, M.~P., McLean, I.~S., Kress, E., \& Wang, E. 2016, https://doi.org/10.1117/12.2233767, 9908, 866, \dodoi{10.1117/12.2233767}

\bibitem[{Martin {et~al.}(2018)Martin, Fitzgerald, McLean, Doppmann, Kassis, Canfield, Johnson, Kress, Aliado, Lanclos, Magnone, Sohn, Wang, \& Weiss}]{Martin2018}
Martin, E.~C., Fitzgerald, M.~P., McLean, I.~S., {et~al.} 2018, https://doi.org/10.1117/12.2312266, 10702, 63, \dodoi{10.1117/12.2312266}

\bibitem[{{M}c{K}inney(2010)}]{mckinney-proc-scipy-2010}
{M}c{K}inney. 2010, in {P}roceedings of the 9th {P}ython in {S}cience {C}onference, ed. {S}t\'efan van~der {W}alt \& {J}arrod {M}illman, 56 -- 61, \dodoi{10.25080/Majora-92bf1922-00a}

\bibitem[{McLean {et~al.}(1998)McLean, Becklin, Bendiksen, Brims, Canfield, Figer, Graham, Hare, Lacayanga, Larkin, Larson, Levenson, Magnone, Teplitz, \& Wong}]{McLean1998}
McLean, I.~S., Becklin, E.~E., Bendiksen, O., {et~al.} 1998, https://doi.org/10.1117/12.317283, 3354, 566, \dodoi{10.1117/12.317283}

\bibitem[{Melin(2020)}]{Melin2020}
Melin, H. 2020, h3ppy,  GitHub.
\newblock \url{https://github.com/henrikmelin/h3ppy}

\bibitem[{Melin {et~al.}(2005)Melin, Miller, Stallard, \& Grodent}]{Melin2005}
Melin, H., Miller, S., Stallard, T., \& Grodent, D. 2005, Icarus, 178, 97, \dodoi{10.1016/J.ICARUS.2005.04.016}

\bibitem[{{Melin} \& {Stallard}(2016)}]{Melin2016}
{Melin}, H., \& {Stallard}, T.~S. 2016, \icarus, 278, 238, \dodoi{10.1016/j.icarus.2016.06.023}

\bibitem[{Melin {et~al.}(2024)Melin, O’Donoghue, Moore, Stallard, Fletcher, Roman, Harkett, King, Thomas, Wang, Tiranti, Knowles, de~Pater, Fouchet, Fry, Wong, Holler, Hueso, James, Orton, Mura, Sánchez-Lavega, Lellouch, de~Kleer, \& Showalter}]{Melin2024}
Melin, H., O’Donoghue, J., Moore, L., {et~al.} 2024, Nature Astronomy 2024 8:8, 8, 1000, \dodoi{10.1038/s41550-024-02305-9}

\bibitem[{Mendillo {et~al.}(2022)Mendillo, Narvaez, Moore, \& Withers}]{Mendillo2022}
Mendillo, M., Narvaez, C., Moore, L., \& Withers, P. 2022, Journal of Geophysical Research: Planets, 127, e2021JE007169, \dodoi{10.1029/2021JE007169}

\bibitem[{Migliorini {et~al.}(2019)Migliorini, Dinelli, Moriconi, Altieri, Adriani, Mura, Connerney, Atreya, Piccioni, Tosi, Sindoni, Grassi, Bolton, Levin, Gérard, Noschese, Cicchetti, Sordini, Olivieri, \& Plainaki}]{Migliorini2019}
Migliorini, A., Dinelli, B.~M., Moriconi, M.~L., {et~al.} 2019, Icarus, 329, 132, \dodoi{10.1016/J.ICARUS.2019.04.003}

\bibitem[{Miller {et~al.}(1997)Miller, Achilleos, Ballester, Lam, Tennyson, Geballe, \& Trafton}]{Miller1997}
Miller, S., Achilleos, N., Ballester, G.~E., {et~al.} 1997, Icarus, 130, 57, \dodoi{10.1006/ICAR.1997.5813}

\bibitem[{Miller {et~al.}(1990)Miller, Joseph, \& Tennyson}]{Miller1990}
Miller, S., Joseph, R.~D., \& Tennyson, J. 1990, The Astrophysical Journal, 55, 55, \dodoi{10.1086/185811}

\bibitem[{Miller {et~al.}(2010)Miller, Stallard, Melin, \& Tennyson}]{Miller2010}
Miller, S., Stallard, T., Melin, H., \& Tennyson, J. 2010, Faraday Discussions, 147, 283, \dodoi{10.1039/C004152C}

\bibitem[{Miller {et~al.}(2020)Miller, Tennyson, Geballe, \& Stallard}]{Miller2020}
Miller, S., Tennyson, J., Geballe, T.~R., \& Stallard, T. 2020, Reviews of Modern Physics, 92, \dodoi{10.1103/RevModPhys.92.035003}

\bibitem[{Moore {et~al.}(2018)Moore, Yadav, Kulowski, Cao, Bloxham, Connerney, Kotsiaros, Jørgensen, Merayo, Stevenson, Bolton, \& Levin}]{Moore2018}
Moore, K.~M., Yadav, R.~K., Kulowski, L., {et~al.} 2018, Nature, 561, 76, \dodoi{10.1038/S41586-018-0468-5}

\bibitem[{{Moore} {et~al.}(2019){Moore}, {Melin}, {O'Donoghue}, {Stallard}, {Moses}, {Galand}, {Miller}, \& {Schmidt}}]{Moore2019}
{Moore}, L., {Melin}, H., {O'Donoghue}, J., {et~al.} 2019, Philosophical Transactions of the Royal Society of London Series A, 377, 20190067, \dodoi{10.1098/rsta.2019.0067}

\bibitem[{M{\"{u}}ller-Wodarg {et~al.}(2019)M{\"{u}}ller-Wodarg, Koskinen, Moore, Serigano, Yelle, H{\"{o}}rst, Waite, \& Mendillo}]{Mueller-Wodarg2019}
M{\"{u}}ller-Wodarg, I.~C., Koskinen, T.~T., Moore, L., {et~al.} 2019, Geophysical Research Letters, 46, 2372, \dodoi{10.1029/2018GL081124}

\bibitem[{{Nakamura} {et~al.}(2022){Nakamura}, {Terada}, {Tao}, {Terada}, {Kasaba}, {Leblanc}, {Kita}, {Nakamizo}, {Yoshikawa}, {Ohtani}, {Tsuchiya}, {Kagitani}, {Sakanoi}, {Murakami}, {Yoshioka}, {Kimura}, {Yamazaki}, \& {Yoshikawa}}]{Nakamura2022}
{Nakamura}, Y., {Terada}, K., {Tao}, C., {et~al.} 2022, Journal of Geophysical Research (Space Physics), 127, e30312, \dodoi{10.1029/2022JA030312}

\bibitem[{Neale {et~al.}(1996)Neale, Miller, \& Tennyson}]{Neale1996}
Neale, L., Miller, S., \& Tennyson, J. 1996, The Astrophysical Journal, 464, 516, \dodoi{10.1086/177341}

\bibitem[{O'Donoghue {et~al.}(2016)O'Donoghue, Moore, Stallard, \& Melin}]{ODonoghue2016}
O'Donoghue, J., Moore, L., Stallard, T.~S., \& Melin, H. 2016, Nature, 536, \dodoi{10.1038/nature18940}

\bibitem[{{Oka}(1981)}]{Oka1981}
{Oka}, T. 1981, Philosophical Transactions of the Royal Society of London Series A, 303, 543, \dodoi{10.1098/rsta.1981.0223}

\bibitem[{O’Donoghue {et~al.}(2021)O’Donoghue, Moore, Bhakyapaibul, Melin, Stallard, Connerney, \& Tao}]{ODonoghue2021}
O’Donoghue, J., Moore, L., Bhakyapaibul, T., {et~al.} 2021, Nature 2021 596:7870, 596, 54, \dodoi{10.1038/s41586-021-03706-w}

\bibitem[{pandas Development~Team(2020)}]{reback2020pandas}
pandas Development~Team, T. 2020, pandas-dev/pandas: Pandas, latest,  Zenodo, \dodoi{10.5281/zenodo.3509134}

\bibitem[{Prato {et~al.}(2002)Prato, Kim, \& McLean}]{Prato2002}
Prato, L., Kim, S.~S., \& McLean, I. 2002, REDSPEC Data Reduction Manual

\bibitem[{Raynaud {et~al.}(2004)Raynaud, Lellouch, Maillard, Gladstone, Waite, Bézard, Drossart, \& Fouchet}]{Raynaud2004}
Raynaud, E., Lellouch, E., Maillard, J.~P., {et~al.} 2004, Icarus, 171, 133, \dodoi{10.1016/J.ICARUS.2004.04.020}

\bibitem[{{Sandel} {et~al.}(1980){Sandel}, {Broadfoot}, \& {Strobel}}]{Sandel1980}
{Sandel}, B.~R., {Broadfoot}, A.~L., \& {Strobel}, D.~F. 1980, \grl, 7, 5, \dodoi{10.1029/GL007i001p00005}

\bibitem[{{Seiff} {et~al.}(1997){Seiff}, {Kirk}, {Knight}, {Young}, {Milos}, {Venkatapathy}, {Mihalov}, {Blanchard}, {Young}, \& {Schubert}}]{Seiff1997}
{Seiff}, A., {Kirk}, D.~B., {Knight}, T.~C.~D., {et~al.} 1997, Science, 276, 102, \dodoi{10.1126/science.276.5309.102}

\bibitem[{Smith \& Aylward(2009)}]{Smith2009}
Smith, C. G.~A., \& Aylward, A.~D. 2009, Annales Geophysicae, 27, 199, \dodoi{https://doi.org/10.5194/angeo-27-199-2009}

\bibitem[{Stallard {et~al.}(2015)Stallard, Melin, Miller, Badman, Baines, Brown, Blake, O'Donoghue, Johnson, Bools, Pilkington, East, \& Fletcher}]{Stallard2015}
Stallard, T.~S., Melin, H., Miller, S., {et~al.} 2015, Journal of Geophysical Research: Space Physics, 120, 6948, \dodoi{10.1002/2015JA021097}

\bibitem[{Stallard {et~al.}(2017)Stallard, Melin, Miller, Moore, O'Donoghue, Connerney, Satoh, West, Thayer, Hsu, \& Johnson}]{Stallard2017}
---. 2017, Geophysical Research Letters, 44, 3000, \dodoi{10.1002/2016GL071956}

\bibitem[{Stallard {et~al.}(2018)Stallard, Burrell, Melin, Fletcher, Miller, Moore, O’Donoghue, Connerney, Satoh, \& Johnson}]{Stallard2018}
Stallard, T.~S., Burrell, A.~G., Melin, H., {et~al.} 2018, Nature Astronomy 2018 2:10, 2, 773, \dodoi{10.1038/s41550-018-0523-z}

\bibitem[{Strobel \& Smith(1973)}]{Strobel1973}
Strobel, D.~F., \& Smith, G.~R. 1973, Journal of the Atmospheric Sciences, 30, 718, \dodoi{10.1175/1520-0469(1973)030}

\bibitem[{Tao {et~al.}(2011)Tao, Badman, \& Fujimoto}]{Tao2011}
Tao, C., Badman, S.~V., \& Fujimoto, M. 2011, Icarus, 213, 581, \dodoi{10.1016/J.ICARUS.2011.04.001}

\bibitem[{{Tao} {et~al.}(2009){Tao}, {Fujiwara}, \& {Kasaba}}]{Tao2009}
{Tao}, C., {Fujiwara}, H., \& {Kasaba}, Y. 2009, Journal of Geophysical Research (Space Physics), 114, A08307, \dodoi{10.1029/2008JA013966}

\bibitem[{Virtanen {et~al.}(2020)Virtanen, Gommers, Oliphant, Haberland, Reddy, Cournapeau, Burovski, Peterson, Weckesser, Bright, {van der Walt}, Brett, Wilson, Millman, Mayorov, Nelson, Jones, Kern, Larson, Carey, Polat, Feng, Moore, {VanderPlas}, Laxalde, Perktold, Cimrman, Henriksen, Quintero, Harris, Archibald, Ribeiro, Pedregosa, {van Mulbregt}, \& {SciPy 1.0 Contributors}}]{2020SciPy-NMeth}
Virtanen, P., Gommers, R., Oliphant, T.~E., {et~al.} 2020, Nature Methods, 17, 261, \dodoi{10.1038/s41592-019-0686-2}

\bibitem[{Yates {et~al.}(2020)Yates, Ray, Achilleos, Witasse, \& Altobelli}]{Yates2020}
Yates, J.~N., Ray, L.~C., Achilleos, N., Witasse, O., \& Altobelli, N. 2020, Journal of Geophysical Research: Space Physics, 125, e2019JA026792, \dodoi{10.1029/2019JA026792}

\bibitem[{Yelle \& Miller(2004)}]{Yelle2004}
Yelle, R.~V., \& Miller, S. 2004, Jupiter's Thermosphere and Ionosphere, Vol.~1 (Cambridge University Press), 185--218

\end{thebibliography}
\bibliographystyle{aasjournal}


\end{document}